\documentclass{aa}
\usepackage{graphics}
\usepackage{epsf}
\usepackage{rotate}
\begin{document}
\mathindent1.0cm
\topmargin-0.5cm

\thesaurus{06 (08.05.3; 08.23.1; 08.16.6)}

\title{The evolution of helium white dwarfs}
\subtitle{III. On the ages of millisecond pulsar systems}

\author{D. Sch\"onberner\inst{1}
   \and T. Driebe\inst{2}
   \and T. Bl\"ocker\inst{2}}

\institute{Astrophysikalisches Institut Potsdam, An der Sternwarte 16,
           D-14482 Potsdam, Germany (deschoenberner@aip.de)
      \and Max-Planck-Institut f\"ur Radioastronomie, Auf dem H\"ugel 69, 
           D-53121 Bonn, Germany \\
                               (driebe@mpifr-bonn.mpg.de, 
                                bloecker@mpifr-bonn.mpg.de)} 
      
\offprints{D. Sch\"onberner}

\date{Received / Accepted}
\titlerunning{The evolution of helium white dwarfs III}
\maketitle      

\begin{abstract}   

   We employed recently computed evolutionary white-dwarf models with helium
   cores, supplemented by heavier models with carbon-oxygen cores, in order
   to investigate the ages of millisecond pulsar systems based on the cooling
   properties of the compact companions. Contrary to the behaviour of more 
   massive white dwarfs, the evolutionary speed  of low-mass white-dwarf 
   models is substantially slowed down by ongoing hydrogen burning. 
   By comparing the cooling ages of
   these models with the spin-down ages of the pulsars for those
   systems for which reasonable information about the compact companions is
   available, we found good correspondence between both ages. Based on these models 
   any revisions concerning the temporal evolution of millisecond pulsars do not 
   appear to be necessary. 
\end{abstract}

\section{Introduction}

  Millisecond pulsars are thought to be components of low-mass binary systems in their final
  stage of evolution: the neutron star which has been spun up by accretion of
  matter from a low-mass evolved companion is now being slowed down by emission of
  magnetic dipole radiation (recycled radio pulsar). The companion, after having  
  transferred most of its envelope mass towards the neutron star, remains as a 
  white dwarf of rather a low mass whose core consists, in the majority of 
  the known cases, of helium. The characteristic age, or so-called 
  {\em spin-down age}, of the (recycled) pulsar depends on the physics how 
  the neutron star's rotational energy  is converted into non-thermal
  emission of electromagnetic energy. On the other hand,
  the white-dwarf age is ruled by the 
  white dwarf's thermo-mechanical structure and the transformation of 
  gravothermal energy content into thermal emission of photons from the surface.  
  Any age determinations of the pulsar and the dwarf component should give the
  same answer, provided our physical understanding of the pulsar's slow-down
  processes and the white dwarf's cooling properties  is correct. 
  
  So far, no general consensus on this matter has been achieved. Under the
  assumption that the cooling properties of low-mass white dwarfs are ruled
  by rather simple laws as is known from evolutionary calculations of more 
  massive white dwarfs with
  carbon-oxygen cores (cf.\ Iben \& Tutukov \cite{IT84}; 
  Koester \& Sch\"onberner \cite{KS86}, Bl\"ocker \cite{BL95}), 
  large age differences between 
  the pulsars and their dwarf companions have been found. 
  In general, the white dwarfs appear to be much younger than the pulsars
  (cf.\ Hansen \& Phinney \cite{HP98b} for a recent, detailed account).
   The best-studied example is the
  \object{PSR J1012+5307} system, for which Lorimer et al.\ (\cite{LFLN95}) 
  determined 7 Gyr for the spin-down age of the pulsar, but only about 0.3 Gyr
  for the white dwarf's age. Note that the usual spin-down age determinations
  are based on the assumption that the initial rotational period after completion 
  of the spin-up by accretion is much smaller then the present one, and
  that the pulsar emits magnetic dipole radiation (braking index $n=3$). 
   A summary of
   the assumptions inherent in the derivation of characteristic or spin-down
   ages of pulsars is given in Hansen \& Phinney (\cite{HP98b}). 
  A discrepant result as found
  for \object{PSR J1012+5307}, if true, would have important consequences for 
  the details of the accretion process and the following spin-down phase (cf.\
  Burderi et al.\ \cite{BKW96}).
  
  A larger sample of millisecond pulsar systems with white-dwarf companions
  has recently been investigated by Hansen \& Phinney (\cite{HP98b}),
  using a grid of low-mass white-dwarf sequences especially computed for this
  purpurse (Hansen \& Phinney \cite{HP98a}). 
  In most cases spin-down and cooling ages appeared to be discrepant
  to various degrees, and the authors were able to constrain the initial spin
  periods and spin-up histories for individual systems, especially also for the
  \object{PSR J1012+5307} system. However, the white-dwarf
  models which this study is based on, are generated from ad-hoc assumed 
  initial configurations. These configurations appear not to be consistent with respect to 
  the thermo-mechanical structures and unprocessed, hydrogen-rich 
  envelopes with what would be adequate for companions in these pulsar binary 
  systems. 

  The early investigations concerning the evolution of helium white dwarfs made
  by Webbink (\cite{W75}) indicated that the final cooling is slowed down
  considerably by ongoing hydrogen burning via the pp cycle. Obviously the
  cooling behaviour of low-mass white dwarfs depends on the size of the still
  unprocessed hydrogen-rich envelope, i.e.\ whether this envelope is massive
  enough as to sustain burning temperatures at its bottom for a long time span.
  The Webbink (\cite{W75}) white-dwarf models are, however, just evolved main sequence 
  stars without any consideration of mass loss. 
  
  Since white-dwarf envelope masses cannot be guessed from first
  principles, they must rather be determined by detailed evolutionary
  calculations. A step in this direction was made by Alberts et al.\ 
  (\cite{ASH96}) and Sarna et al.\ (\cite{SAAM98}) who modelled 
  the \object{PSR J1012+5307} system and in particular the evolution of the
  mass giving companion. 
  It turned  out that the donor shrinks below
  its Roche lobe while still having a rather massive hydrogen-rich envelope
  which is able to keep hydrogen burning dominant even through the white-dwarf
  cooling phase. The evolution was slowed down
  to such an extent that the discrepancy with the spin-down age of the pulsar 
  vanished completely.
  
  Strictly speaking the strength of hydrogen burning, and hence the cooling age 
  of an observed white dwarf, depends on the size of the envelope before entering 
  the cooling path. This envelope mass can be reduced because of thermal 
  instabilities of the burning shell  
  when the CNO rate dies out, namely by 
\begin{itemize}
       \item
   enhanced hydrogen consumption during the instability (flash) itself, and by
       \item
   a possible Roche-lobe overflow driven by the rapid envelope expansion.
\end{itemize}
  The latter case was dominant for the evolution of the Iben \& Tutukov 
  (\cite{IT86}) 0.3\,M$_{\sun}$ helium white-dwarf model: 
  Roche-lobe overflow due to the flash-driven 
  envelope expansions reduced the envelope mass {\em below\/} the critical 
  value necessary for hydrogen burning. The white-dwarf models of Webbink (\cite{W75})
  and Sarna et al.\ (\cite{SAAM98}) experienced phases of unstable
  hydrogen burning for \mbox{$ M \la 0.2$~M$_{\sun}$} (but see Driebe et al.\ \cite{DBSH99} 
  for a discussion).
  
  Recently Driebe et al.\ (\cite{DSBH98}) published a grid of evolutionary
  tracks for helium white-dwarf models which were generated 
  by enhanced mass loss applied at different positions along the red-giant 
  branch of a 1\,M$_{\sun}$ sequence (see also Iben \& Tutukov \cite{IT86}, Castellani et al.\ 1994). 
  This method mimicks to some extent the
  mass transfer in  binary systems and allows to get reliable post-red-giant
  configurations which are very useful for the interpretation of observations.
  Driebe et al.\ (\cite{DSBH98}) covered the whole 
  mass range of interest, and they demonstrated that
\begin{itemize}
       \item     
    the anti-correlation between core mass and size of envelope (cf.\ Bl\"ocker et al.\ \cite{BHDBS97}) 
    determines later the nuclear activity along the cooling branch, and that 
       \item
    thermal instabilities of the hydrogen-burning shell appear to be restricted 
    to the mass range of approximately 0.2 to 0.3~M$_{\sun}$. 
\end{itemize}  
  The absence of thermal flashes
  below $ M= 0.2$~M$_{\sun}$ agrees well with the results of Alberts et al.\
  (\cite{ASH96}) 
  but disagrees with those of Sarna et al.\ (\cite{SAAM98}). Nevertheless,
  the cooling times of our models are in excellent agreement with both studies. 
  From the
  given parameters of the white-dwarf component in the \object{PSR J1012+5307} 
  system, Driebe et al.\  (\cite{DSBH98}) determined then its age to be of $ 6 \pm 1$~Gyr, in 
  good agreement with the pulsar's spin-down age of $ 7.0 \pm 1.4$~Gyr
  (Lorimer et al.\ \cite{LFLN95}).
  
  The latest effort in a better understanding of the combined pulsar-white dwarf
  systems is that of Burderi et al.\ (\cite{BKW98}). They took the pulsar 
  spin-down ages at their face value and 
  concluded that the standard assumption for the white-dwarf cooling
  (i.e.\ without nuclear burning) complies with the observations, 
  except for masses below approx.\ 0.2~M$_{\sun}$. There are, however, some facts 
  that we would like to point out:
  Burderi et al.\  (\cite{BKW98}) used data 'renormalized' to a standard luminosity of 
  $ 10^{-2}$~L$_{\sun}$, whereby it remains unclear how ages can be renormalized
  if the temporal evolution of the systems is not known a priori. Furthermore,
  they extrapolated existing white-dwarf cooling models into mass regimes where
  they are not valid anymore. 
       
  Because of its importance 
  we  felt the necessity to reconsider the whole issue by utilizing 
  more realistic evolutionary models for low-mass white dwarfs. We will show in the
  next section that with such models a consistent description of those 
  millisecond pulsar binary systems can be achieved for which sufficiently 
  accurate data is available. 
\section{Pulsar characteristic times and cooling ages of their white-dwarf companions}   
\begin{figure*}[th]
\epsfxsize=12cm
\rotate[r]{
\epsfbox{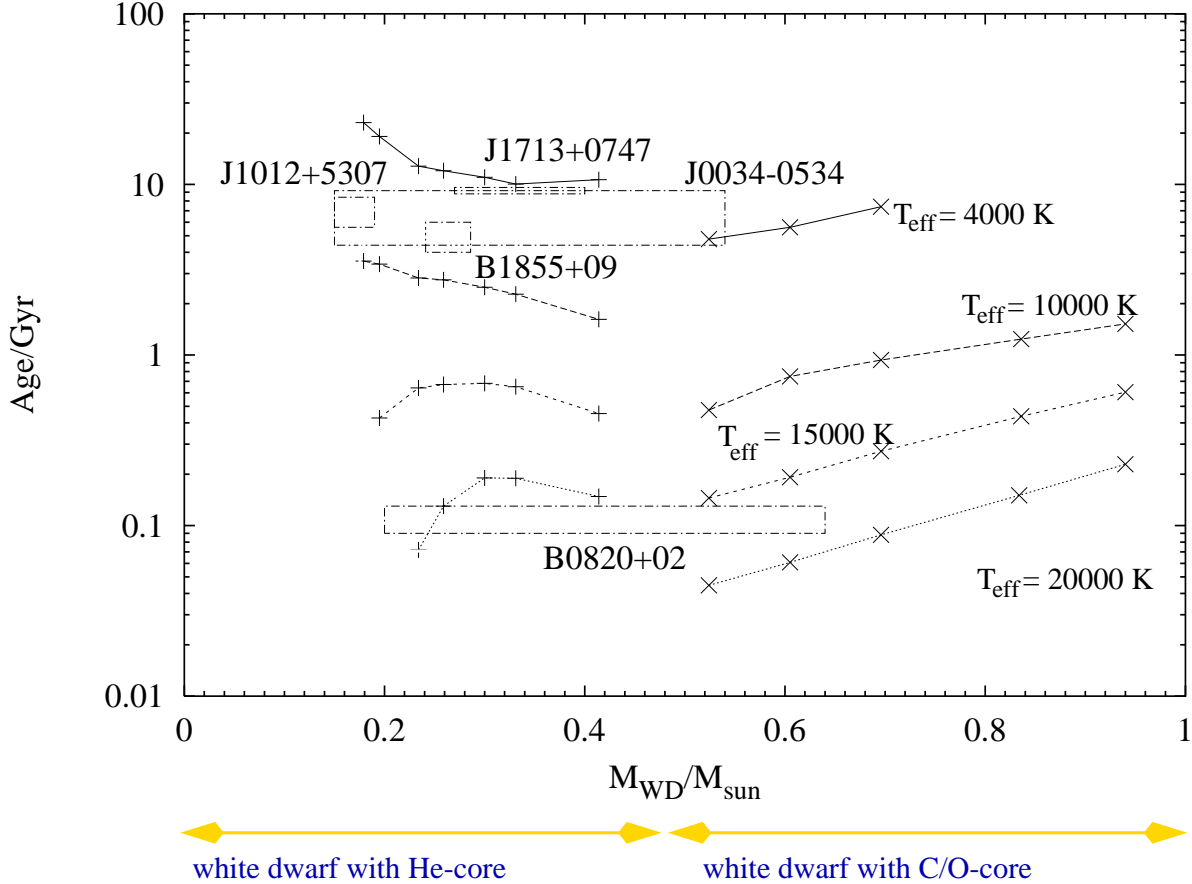}
}
\caption{\label{FIG1} 
              Time-scales vs.\ white-dwarf masses. Boxes indicate millisecond
              pulsar systems with reasonably known spin-down ages and companion
              masses. The sizes of the boxes indicate
              the errors listed in Table 1. The lines with symbols (+: helium
              white dwarfs, $\times$: carbon/oxygen white dwarfs)
              mark the post red-giant ages of
              white-dwarf models at four different temperatures along their
              cooling tracks taken from
              Driebe et al.\ (1998) and Bl\"ocker (\cite{BL95}). For more 
              details see text.}
\end{figure*}

  We  started with the sample of millisecond pulsar systems used by
  Burderi et al.\ (\cite{BKW98}, see their Table 1 and our Fig.~\ref{FIG1}), 
  but made a few changes:
  the companion mass for \object{PSR J1012+5307} was updated
  according to Driebe et al.\ (\cite{DSBH98}), and the systems 
  \object{PSR J1640+2224} and \object{PSR J0437-4715} were omitted because of
  too uncertain pulsar ages. 
  For convenience, the relevant data are collected 
  in Table \ref{TAB1}, and all the listed systems are shown in Fig.\,\ref{FIG1}
  where the characteristic ages of the pulsars are plotted against the 
  possible mass ranges of their (white-dwarf) companions. 

The last column in Table \ref{TAB1} gives the white-dwarf masses
according to the binary evolution calculations of Tauris \& Savonije (\cite{TS99}).
Within these binary calculations a relation between the system's orbital period  $P_{\rm orb}$
and the white-dwarf mass can be derived (see e.\ g.\  Savonije \cite{S87} and
Rappaport et al.\ \cite{RPJSH95}). For the systems discussed here the
$P_{\rm orb}-M_{\rm WD}$ relation of Tauris \& Savonije (\cite{TS99}) predicts
masses which are well within the estimated mass limits (see Table \ref{TAB1}).

   It should be emphasized that the characteristic ages given in Table \ref{TAB1}
   and plotted in Fig.\,\ref{FIG1} are based on certain assumptions (see
   Introduction) which may
   not be fulfilled in all cases. The corresponding systematic errors are 
   difficult to assess and cannot be accounted for in this study. 

  Also shown in  Fig.\,\ref{FIG1} are (post-red giant) ages of helium 
  white-dwarf models taken from Driebe et al.\ (\cite{DSBH98}), supplemented by
  ages from evolutionary white-dwarf models with carbon-oxygen cores 
  (Bl\"ocker \cite{BL95}). The ages are given for four effective
  temperatures as to simplify the comparisons with the observed sytems: the
  range between 4\,000 and 20\,000 K embraces roughly the estimated effective
  temperatures of the companions of the systems \object{PSR J0034-0534},
  \object{PSR J1713+0747}, \object{PSR J1012+5307} and  \object{PSR B0820+02} 
  (cf.\ Hansen \& Phinney \cite{HP98b}).  Only for the \object{PSR J1012+5307} companion exists 
  a rather accurately  determined effective temperature (8\,600~K, van Kerkwijk
  et al.\ \cite{KBK96}, Callanan et al.\ \cite{CGK98}). 
 
  Note that the model ages are counted from the beginning of the post red-giant 
  phase, i.e.\ they include also the contraction towards the white-dwarf
  regime.
  For white dwarfs of low mass this contraction from a giant towards
  a white-dwarf configuration makes up for a significant fraction of their ages 
  and must be accounted for in younger systems like \object{PSR B0820+02}.   
\begin{table}[tb] 
  \caption{Mass estimates for the white dwarfs and characteristic
           ages $\tau_{\rm c}$ of the pulsars. References are listed in the fourth column.
           The fifth column gives the white-dwarf mass from the $P_{\rm orb}-M_{\rm WD}$
           relation of Tauris \& Savonije (\cite{TS99}).}   \label{TAB1} 
  \begin{tabular}{lcccc}
    \hline\noalign{\smallskip}
    PSR          &    Mass limits     &  $\tau_{\rm c}$  &   Ref.   &  $M_{\rm WD}^{\rm TS99}$\\
                 &    [M$_{\sun}$]    &      [Gyr]       &          &         M$_{\sun}$]    \\
    \noalign{\smallskip}
    \hline\noalign{\smallskip}          
    J1012+5307   &     0.15$-$0.19    &   $ 7.0\pm 1.4  $   & 1    & 0.19\\
    B0820+02     &     0.20$-$0.64    &   $ 0.11\pm0.02 $   & 1    & 0.50\\
    B1855+09     &     0.24$-$0.29    &   $ 5.0\pm 1.0  $   & 1,2  & 0.26\\
    J0034$-$0534 &     0.15$-$0.54    &   $ 6.8\pm 2.4  $   & 3,4  & 0.21\\
    J1713+0747   &     0.27$-$0.40    &   $ 9.2\pm 0.4  $   & 3,5  & 0.33\\
    \noalign{\smallskip}\hline \noalign{\smallskip}
  \end{tabular}
   1: Burderi et al.\ (\cite{BKW98}) \\
   2: van Kerkwijk et al.\ (\cite{Kerk}) \\
   3: Hansen \& Phinney (\cite{HP98b})\\
   4: van Kerkwijk (priv. comm.)\\
   5: Camilo et al.\ (\cite{CFW94})
\end {table}
 
The temporal behavior of the models, as shown in Fig.~\ref{FIG1}, is 
determined by the following facts: 
\begin{itemize}

  \item  
  The compositional differences between the lighter and heavier white dwarfs 
  causes the obvious age break around $M_{\rm WD}=0.5\, {\rm M}_{\sun}$.

  \item 
  Hydrogen burning in the helium white dwarfs is, for a given temperature, 
  responsible for the strong increase in age with decreasing mass. 

  \item
   At high temperatures, the lighter models
  ($\la 0.23$~M$_{\sun}$) are still in a pre white-dwarf phase very close to 
  the turn-around point with rather low (post red-giant) ages. 
\end{itemize} 

    Fig.\,\ref{FIG1} clearly demonstrates that for an effective temperature of
    4\,000~K our low-mass white-dwarf models exceed ages of 10~Gyr, 
    roughly consistent with the pulsar characteristic ages. From the positions
    of individual pulsars (with error bars) we can estimate effective 
    temperature and gravity ranges, and also the internal composition, to be 
    expected for the white-dwarf companions. The results are collected in 
    Table \ref{TAB2}. Since for some of the white dwarfs temperature estimates
    based on photometry are available (Hansen \& Phinney \cite{HP98b}), 
    consistency checks are possible. 

\begin{table}[tb] 
  \caption{Estimates of effective temperature and surface gravity 
           for the white-dwarfs in the MSP systems from Table \ref{TAB1}.
           The fourth column gives the core composition of the white dwarf.
  }
  \label{TAB2} 
  \begin{tabular}{lcccc}
    \hline\noalign{\smallskip}
    PSR          &   $T_{\rm eff}$ [K]    &     $\log g$ & core & Ref.\\
                 &                        &              &      &     \\
    \noalign{\smallskip}
    \hline\noalign{\smallskip}       
    J1012+5307   &  $8\,550\pm 25$        &   $6.75\pm0.07$   & He & 1\\
                 &  $8\,670\pm300$        &   $6.34\pm0.20$   & He & 2\\
    B0820+02     &  $20\,000\dots22\,000$ &   $6.0\dots7.4$   & He & 3\\
                 &  $15\,000\dots18\,000$ &   $7.75\dots8.0$  & C/O& 3\\
    B1855+09     &  $7\,000\dots9\,000$   &   $7.2\pm0.1$ & He & 3\\
    J0034$-$0534 &  $5\,000\dots8\,500$   &   $7.2\pm0.5$ & He & 3\\
                 &  $\la 4000$      &   $\approx7.8$ & C/O & 3\\
    J1713+0747   &  $4\,000\dots4\,500$   &   $7.35\pm0.05$ & He & 3\\
    \noalign{\smallskip}\hline \noalign{\smallskip}
  \end{tabular}

   1: van Kerkwijk et al.\ (\cite{KBK96}) \\
   2: Callanan et al.\ (\cite{CGK98}) \\
   3: present work\\
\end {table}     
  
\subsection{\object{PSR J1012+5307}}

\begin{figure}
\epsfxsize=8.8cm
\epsfbox{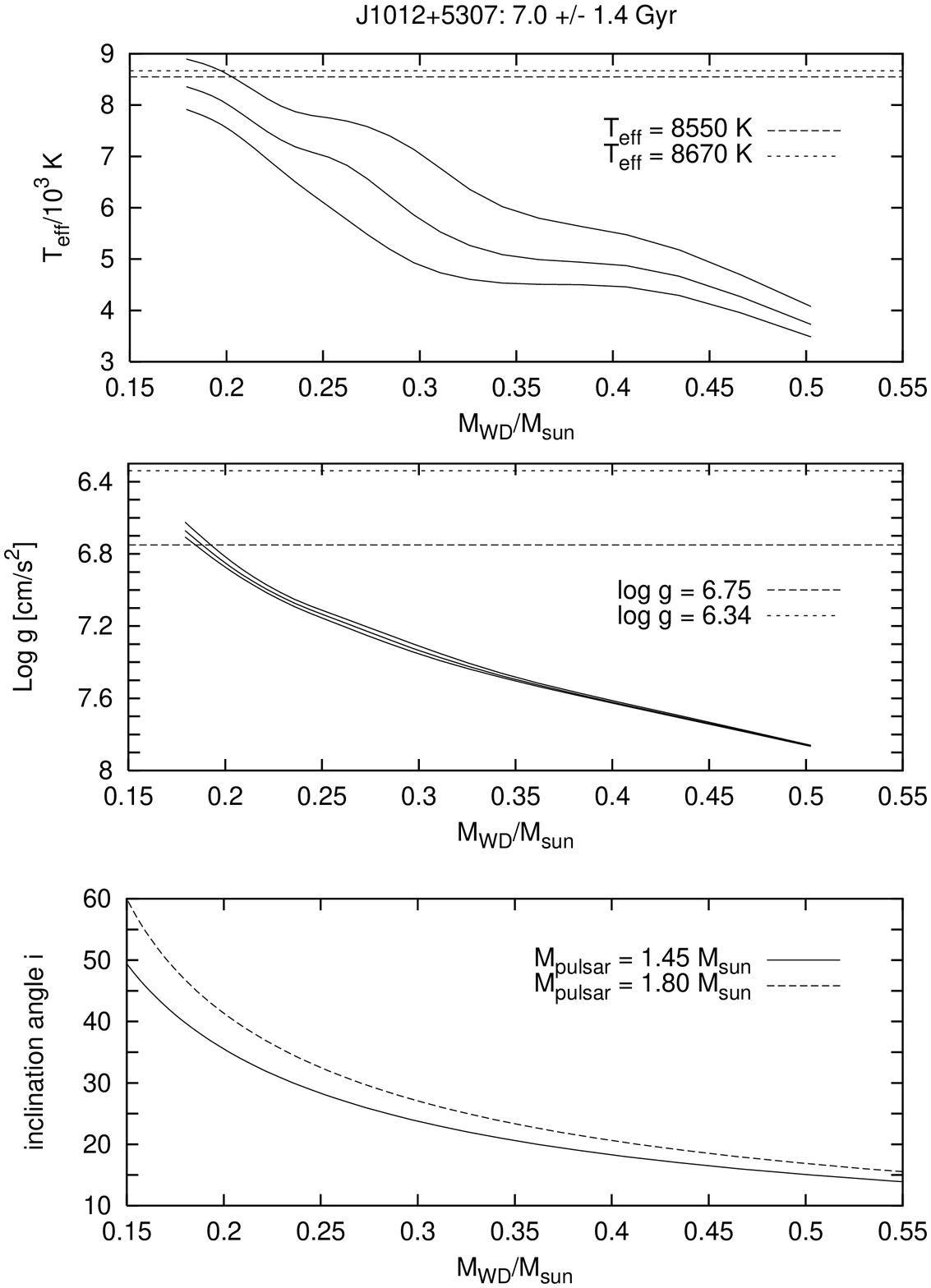}
\caption{\label{FIG2}
            {\bf Upper panel:} Isochrones for three white-dwarf ages, 
            $7.0 \pm 1.4$~Gyr, in an effective temperature vs.\ mass diagram. 
            The effective temperature of the  white-dwarf companion of 
            \object{PSR J1012+5307} is indicated by the horizontal dashed lines
            (long dashed: van Kerkwijk et al.\,\cite{KBK96}; short dashed: 
            Callanan et al.\,\cite{CGK98}).
            {\bf Middle panel:} The same isochrones but in a surface
            gravity vs.\ mass diagram. Again the observed values for
            \object{PSR J1012+5307} are also given (dashed lines). 
            {\bf Lower panel:} The \object{PSR J1012+5307}
            system's inclination for two pulsar masses vs.\ the companion mass.
          }     
\end{figure}

  The white dwarf in  \object{PSR J1012+5307} is so far the best studied 
  companion of all known systems (van Kerkwijk et al.\ \cite{KBK96}; 
  Callanan et al.\ \cite{CGK98}). With its surface parameters known, and 
  together with our evolutionary white-dwarf models, a consistent description 
  of the whole system is found  (Fig.\,\ref{FIG2}, upper two panels). Since the
  mass ratio of both components is known, the system's inclination can be
  determined as well. The course of the inclination angle with white-dwarf 
  companion mass is shown in the lower panel of Fig.\,\ref{FIG2} 
  for two limiting pulsar masses. The inclination can be expected to lie
  between 40 and 50 degrees.

\begin{table*}[p] 
  \caption{ 
  Data for selected millisecond pulsars.
  $P_{\rm spin}$: pulsar's spin period; $\dot{P}$: change of $P_{\rm spin}$;
  $\tau_{\rm c}$: characteristic age; $f(M)$: mass function;
  $M_{\rm WD}^{\rm TS99}$: white-dwarf mass according to the $P_{\rm orb}-M_{\rm WD}$ 
  of Tauris \& Savonije (\cite{TS99}); $T_{\rm eff}, g$: 
  predicted effective temperature and surface gravity for $M_{\rm WD}=M_{\rm WD}^{\rm TS99}$
  (compare Fig.\ \ref{FIG4}).
   }
  \label{TAB3} 
  \begin{tabular}{lcccccccc}
    \hline\noalign{\smallskip}
    PSR  &  $P_{\rm spin}$ & $\dot{P}$    & $\tau_{\rm c}$   & $f(M)$  
         & Ref. & $M_{\rm WD}^{\rm TS99}$ & $T_{\rm eff}$ & $\log g$\\
         &  [ms]           & [$\rm 10^{-20}\, ss^{-1}$] & [Gyr]    
         & [$10^{-3}{\rm M}_\odot$]    &    &  [${\rm M}_\odot$] & [K] & [${\rm cm s}^{-2}$]\\
    \noalign{\smallskip}
    \hline\noalign{\smallskip}       
    \object{B1953+29}    & 6.1332 & 2.95 & 3.3 &  2.417 & 1 & 0.35 &$ 8500^{+3500}_{-1500}$& 7.4\\
    \object{J0751+1807}  & 3.4788 & 0.8  & 7.3 &  0.967 & 2 & 0.18 &$ 7600^{+500}_{-500} $& 6.8\\
    \object{J1045-4509}  & 7.4742 & 1.9  & 6.3 &  1.765 & 1 & 0.23 &$ 7300^{+1000}_{-800} $& 7.1\\
    \object{J1640+2224}  & 3.1633 & 0.29 & 17  &  5.907 & 1 & 0.37 &$ 4000^{+800}_{-500} $& 7.6\\
    \object{J1643-1224}  & 4.6216 & 3.3  & 2.3 &  0.783 & 1 & 0.36 &$10000^{+8000}_{-3000}$& 7.4\\
    \object{J1804-2718}  & 9.3430 & 4.2  & 3.5 &  3.347 & 3 & 0.26 &$ 9000^{+3000}_{-1000}$& 7.1\\
    \object{J1911-1114}  & 3.6257 & 1.34 & 4.4 &  0.797 & 3 & 0.22 &$ 9000^{+2000}_{-1000}$& 7.0\\ 
    \object{J0437-4715}  & 5.7575 & 5.73 & $\le$ 6 & 1.239 & 4 & 0.240 &$ 9500^{+4000}_{-2000}$& 7.0\\
    \object{J2129-5721}  & 3.7263 & 2.0  & 3.0 &  1.049 & 3 & 0.244 &$ 9000^{+4500}_{-1000}$& 7.0\\
    \noalign{\smallskip}
    \hline\noalign{\smallskip} 
  \end{tabular}\\
   1: Taylor et al.\ (\cite{TML93}, 1995 (unpubl.\ work, available 
      via anonymous ftp from {\tt pulsar.princeton.edu} (128.112.84.73))) \\
   2: Lundgren et al.\ (\cite{LZC96b}) \\
   3: Lorimer et al.\ (\cite{LOETAL96})\\
   4: Sandhu et al.\ (\cite{SAETAL97})
\end{table*}    
\begin{figure*}[h]
\epsfxsize=12cm
\rotate[r]{
\epsfbox{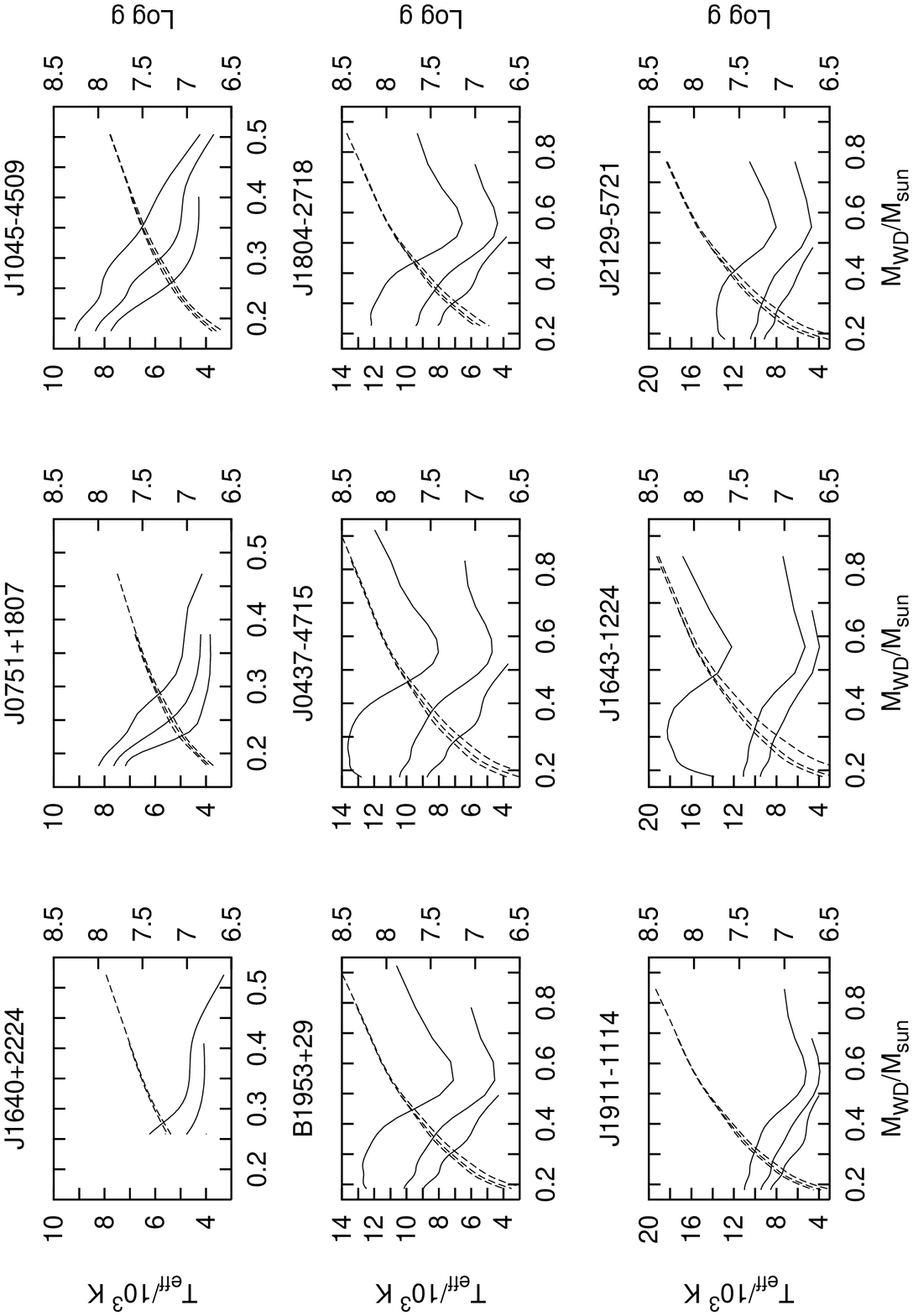}
}
\caption{\label{FIG4}
            Possible effective temperatures (left y-axis, solid lines) 
            and surface gravities (right y-axis, dashed lines)
            as a function of white-dwarf mass for the millisecond pulsar systems of
            Table \ref{TAB3} for three different cooling ages. 
            The middle lines belong to the given characteristic
            age (see Table \ref{TAB3}), except for \object{J1640+2224} where
            the white-dwarf properties are shown for $\tau = 10$~Gyr. The upper and lower
            lines refer to ages $\tau_{\rm c}\pm2$ Gyr. For \object{J0437-4715}, where
            only an upper limit for $\tau_{\rm c}$ is known,
            lines are given for $\tau=1, 3$ and 6 Gyr.
            Note that the lower age limit is represented
            by the upper lines for $T_{\rm eff}$ and the lower lines for $g$.
            }     
\end{figure*}

\subsection{\object{PSR B0820+02}}


   The mass of the companion is rather ill-defined, and it could have a helium
   or carbon/oxygen core (cf.\ Table~\ref{TAB2}). However, the temperature 
   estimate, $ 15\,250 \pm 250$~K (Hansen \& Phinney \cite{HP98b}), allows only a C/O
   white dwarf with at least $\sim$ 0.5~M$_{\odot}$ and a rather high gravity ($\log g\approx 8$),
   in agreement with the results of Tauris \& Savonije (\cite{TS99}) (cf.\ Table \ref{TAB1}).



\subsection{\object{PSR B1855+09}}

       
   The temperature of the companion has recently been photometrically determined
   by van Kerkwijk et al.\ (\cite{Kerk}) to be
   $ T_{\rm eff} = 4\,800 \pm 800$~K. Given its accurately known mass of 
   $0.258^{+0.028}_{-0.016}$~M$_{\odot}$ due to the measured Shapiro delay of pulsar timing
   (Kaspi et al.\ \cite{KTR94}),
   this low temperature corresponds to a cooling age
   of 10~Gyr using our models and 3~Gyr using the models of Hansen \& Phinney 
   (\cite{HP98a}) with their smaller envelope masses ($\le 3 \times 
   10^{-4}$~M$_{\odot}$). 
   The characteristic age of the pulsar is, however, 5~Gyr (cf.\ Table \ref{TAB1}),
   a fact which on one hand might be a hint for a smaller braking index of the pulsar
   as discussed by van Kerkwijk et al.\, (\cite{Kerk}). On the other hand this result 
   strengthens the dependence of age on the thickness of the hydrogen envelope.
    
   According to our 0.259~M$_{\odot}$ model, 
   hydrogen burning ceases at about 10~Gyr, leaving a final 
   unprocessed envelope
   \footnote{The mass of the chemically homogeneous envelope is defined as
   the total mass of the layers above the hydrogen exhausted core.}
   of $ 5\cdot 10^{-4}$~M$_{\odot}$. 
   The envelope mass right after the last shell flash is  
   $\approx 2\cdot 10^{-3}$~M$_{\odot}$, a value which is, however, subject to 
   uncertainties like flash strength and metallicity. In additional calculations 
   we artificially
   reduced this envelope mass by invoking mass loss on the upper cooling branch
   and followed the evolution of the models in the usual manner.
   It turned out that an
   earlier reduction of the envelope mass to $\approx 5\cdot 10^{-4}\, {\rm M}_\odot$
   after $\approx 0.5$ Gyr (at $T_{\rm eff}\approx 10^4\,{\rm K}$)
   would be sufficient to give a cooling age of 5~Gyr at the desired effective temperature
   of \hbox{4800 K.} We note that with this reduced envelope mass, hydrogen burning becomes
   insignificant below $T_{\rm eff}\approx 10^4\,{\rm K}$.

      
\subsection{\object{PSR J0034-0534}}

   Hansen \& Phinney (\cite{HP98b}) estimated a very low temperature limit of
   $ T_{\rm eff} < 3\,500$~K for the white dwarf, yielding ages of more
   than 10~Gyr if helium models are used, which is to be
   compared with the pulsar's characteristic age of $ 6.8\pm2.4$~Gyr (Fig.\,\ref{FIG1}). 
   Consistency between the pulsar's and the white dwarf's age can only be
   achieved if we assume the white dwarf to have a carbon/oxygen core and a mass of $\approx 0.5$~M$_{\sun}$.
   Such a model cools considerably faster and reaches 3\,500~K well within about 6~Gyr.
   We note that this mass is noticeably larger than estimates of other studies. The 
   $P_{\rm orb}-M_{\rm WD}$ relation of Tauris \& Savonije (\cite{TS99}) gives $M\approx 0.21\,{\rm M}_\odot$,
   and the photometric measurements of Lundgren et al.\ (\cite{LFC96a}) $M\approx 0.23\,{\rm M}_\odot$.
   The cooling models of Hansen \& Phinney (\cite{HP98b}) predict $M_{\rm WD}\approx 0.32\,{\rm M}_\odot$
   as an upper limit.
   The discrepancy between these results and our mass
   estimate (C/O-white dwarf) might be related to the same problem as in the case of
   \object{PSR B1855+09}. If the companion is a helium-white dwarf with $M\approx
   0.2\dots0.3\,{\rm M}_\odot$ it is prone to hydrogen shell flashes with the corresponding uncertainties.



\subsection{\object{PSR 1713+0747}}

   Hansen \& Phinney (\cite{HP98b}) give a white-dwarf effective temperature
   of $ 3\,400 \pm 300$~K which is just slightly lower than the one
   predicted by our helium models (see Table \ref{TAB2}). The white-dwarf mass is
   not very sensitive to the temperature value and consistent with the estimate
   of $M_{\rm WD}\approx 0.33\, {\rm M}_\odot$ from Tauris \& Savonije (\cite{TS99}).

\subsection{Other millisecond pulsar systems}

In addition to the sample discussed above we investigated some other 
MSP systems  with respect to the possible $(g,T_{\rm eff})$ 
combinations for the white-dwarf components (see Table \ref{TAB3}). The results
are illustrated in Fig.\,\ref{FIG4} where effective temperature and surface
gravity are plotted as a function of the white-dwarf mass for the systems
listed in Table \ref{TAB3}.
The white-dwarf masses can be estimated from the $P_{\rm orb}-M_{\rm WD}$ relation 
of Tauris \& Savonije (\cite{TS99}). Taking these masses, it is straightforward to
determine effective temperatures and surface gravities of the white-dwarf companions
(see Table \ref{TAB3}).

 
We find the systems in the first row of Fig.\ \ref{FIG4} (J1640+2224, J0751+1807, J1045-4509) 
to be consistent with He-WD companions if $ T_{\rm eff} \ga 4\,000$~K. At this temperature
our helium models reach cooling ages comparable with the age of the galactic disk ($\approx 10$ Gyr).
For the white dwarf in \object{J1640+2224} a temperature estimate is available: 
$ 4\,500\pm1100$~K (Hansen \& Phinney \cite{HP98b}). This value implies, for 10~Gyr, 
a white-dwarf mass of $ \approx 0.35\pm0.05$~M$_{\sun}$ with a surface gravity of 
$ \log g \approx 7.6$ (cf.\ Fig.\,\ref{FIG4}).
This mass is in good agreement with the result of Tauris \& Savonije (\cite{TS99})
($M_{\rm WD}^{\rm TS99}\approx 0.37\, {\rm M}_\odot$). 

For the other systems the temperature ambiguity
(more than one white dwarf mass for a given temperature, see.\ Fig.\ \ref{FIG4}) 
does not allow to exclude
the companions to be C/O-white dwarfs, but the mass values of Tauris \& Savonije (\cite{TS99})
indicate that all these systems should contain helium white dwarfs.


\section{Conclusions}
 
   From the present paper, together with previous efforts which concentrated 
   solely on the \object{PSR J1012+5307} system (Alberts et al.\ \cite{ASH96}; Sarna
   et al.\ \cite{SAAM98}; Driebe et al.\ \cite{DSBH98}), it becomes obvious
   that a consistent description of the millisecond binary systems with
   compact companions can only be achieved by using 
   evolutionary model calculations of white dwarfs which include their complete
   pre-white-dwarf history. 
   The key to the solution of the apparent age
   paradoxon between the pulsar and its white-dwarf companion is the fact that 
   low-mass white dwarfs have massive, still unburnt envelopes that sustain
   hydrogen burning at their bases for a long time. Hydrogen burning slows 
   down the cooling of a low-mass white dwarf to such an extent that cooling 
   ages become comparable to, or may even exceed, observed pulsar 
   spin-down ages. 
   
   Employing our evolutionary helium white-dwarf models, supplemented by those
   with carbon-oxygen cores, we demonstrated that, next to the already studied
   PSR J1012+5307 system, also in other millisecond pulsar binary 
   systems with reliable information on pulsar age and companion
   properties, as mass and temperature, reasonable agreement between the
   components' ages is achieved.
   The use of white-dwarf models with ad hoc assumed envelope masses 
   may lead to erroneous interpretations 
   since these envelope masses are usually much smaller than those which
   follow from complete evolutionary calculations. 

   It appears to us that upon
   using  realistic white-dwarf models in interpreting
   millisecond binary systems there is no need to modify existing ideas of
   the spin-down process of pulsars. Clearly a larger sample of 
   well-studied systems like \object{PSR J1012+5307} would be very important
   in investigating more precisely the cooling theory of white dwarfs {\it and\/}
   the braking of radio pulsars. 

\begin{acknowledgements}
We would like to thank Marten van Kerkwijk and Gerrit Savonije for helpful discussions
and comments.
\end{acknowledgements}

\end{document}